\documentclass[useAMS,usenatbib,usegraphicx]{mn2e}
\usepackage{amsmath,amssymb,amsbsy,amsfonts}
\usepackage{amssymb}
\usepackage[british]{babel}
\usepackage{mdwmath}
\usepackage{amsmath}
\usepackage{color}
\usepackage{url}
\usepackage{rotating}
\title[Understanding the light curves of HST-1 knot in M87 as shock waves]
{Understanding the light curves of the HST-1 knot in M87 with internal relativistic shock waves along its jet}
\author[Y. Coronado, O. L\'opez-Corona \& S. Mendoza]{Y.
Coronado$^1$,
 O. L\'opez-Corona$^{1,2}$ \& S. Mendoza$^1$\thanks{E-mail address: 
\{coronado,olopez,sergio\}@astro.unam.mx.} \\
\(^{1}\) Instituto de Astronom\'{\i}a, Universidad Nacional Aut\'onoma
de M\'exico, AP  70-264, Distrito Federal 04510, M\'exico\\
$^{2}$Centro de Ciencias de la Complejidad, Universidad Nacional
Aut\'onoma de M\'exico, Ciudad Universitaria, Distrito Federal, 
M\'exico.}

\begin{document}

\date{}


\maketitle

\label{firstpage}

\begin{abstract}
 Knots  or blobs observed in astrophysical jets are commonly
interpreted as shock waves moving along them.  Long time observations
of the HST-1 knot inside the jet of the galaxy M87 have produced
detailed multi-wavelength light curves.  In this article, we model
these light curves using the semi-analytical approach developed by
\citet{mendoza09}. This model was developed to account for the light
curves of working surfaces moving along relativistic jets. These working
surfaces are generated by periodic oscillations of the injected flow
velocity and mass ejection rates at the base of the jet.   Using genetic
algorithms to fit the parameters of the model, we are able to explain
the outbursts observed in the light curves of the HST-1 knot with an
accuracy greater than a 2-\(\sigma\) statistical confidence level.
\end{abstract}


\section{Introduction}
\label{introduction}

  The jet in the galaxy M87 was detected in the optical band by
\citet{Curtis18}.   It is the closest Active Galaxy Nuclei with a redshift
\( z = 0.004360 \) and has been extensively monitored in multi-frequency
campaigns, particularly over the last decade.  Radio interferometry
and high resolution optical and X-ray observations show the complex
structures formed inside the jet as close as \(\sim 100 \text{pc} \)
from the nucleus \citep{waters05}. The most exotic of these structures,
is a particular knot formed in 1999 and labelled HST-1. The evolution
of HST-1 began to be closely followed in 2000 with the Chandra X-ray
telescope \citep{harris03,harris06, harris09} since  it started to develop
a rapid increase on its X-ray emission, achieving a maximum in 2005,
corresponding to a factor of 50 as compared to the emission detected
in 2000.  After this maximum, the emission decreases and is followed by
two further increments in 2006 and 2008.  Ultraviolet \citep{madrid09}
and radio \citep{Chang10} observations show a similar behaviour of its
light curve.  The whole emission of M87 presents an optical outburst in
2005 \citep{madrid09} which is related to the maximum emission of the
HST-1 knot in the same year.   This strongly suggests that the outburst
is produced by the strong emission of the knot.

  Knots in astrophysical jets are usually identified with internal shock
waves travelling along the jet.  These internal shock waves can be produced by
different mechanisms: (a) interactions of the jet with an overdense medium,
e.g. clouds \citep[cf.][]{mendoza-phd,mendoza-1-01}, (b) bending of jets
above a critical value \citep{mendoza-phd,mendoza-1-02}, and (c) Periodic
variations of the injected velocity and mass at the base of the jet
\citep[e.g.][and references therein]{rees94,jamil08,mendoza09}.   

  In the literature, the main contribution of the X-ray emission of the
HST-1 knot  is still under
discussion and the interpretations vary between an effect of a hot
accretion disc with the corona \citep{marscher02} and a particular
phenomena of a re-collimation shock \citep{Stawarz06}, causing the
impressive flare in X-rays. Later observations in radio revealed
superluminal motions in HST-1 being a well isolated knot from the nucleus
\citep{biretta99},
displaced from the central engine by $\geq 120 \ \text{pc}$ \citep{cheung07}. All
this makes HST-1 the best studied knot for a possible internal shock
mechanism inside a jet. It is also an ideal target to observe due to its
proximity. The strong multi-wavelength emission from the jet and its knots 
allow to test the physics of knots
and shock waves in the relativistic regime.

  Since relativistic outbursts are usually thought of as internal shock
waves travelling along the jet, produced by periodic variations of the
injected flow, it is quite natural to model the high emission light
curve of the HST-1 knot as shock waves produced by this mechanism.
The semi-analytical model by \citet{mendoza09} (denoted as M09 in
what follows) has been quite useful in modelling not only outbursts
associated to long gamma-ray bursts  but also to the many outbursts
detected on the light curve of the blazar PKS~1510-089 \citep{cabrera13}.
We show in this article, that such a model is also good for modelling
and understanding the multi-frequency features observed in the HST-1 knot
of the M87 galaxy.

  \citet{harris09} found a quasi-periodic impulse signature in
the brightening and dimming of the core of M87.  This was interpreted as
a manifestation of past modulation of jet power, possibly  by a local
oscillation of the process that converts the bulk kinetic jet power to
the internal energy of the emitting plasma.  This result reinforces the
use of the M09 model in order to explain the formation and evolution of
the HST-1 knot.

  The  article is organised as follows. In section~\ref{observations}
we present the multi-wavelength observation campaigns of the HST-1 knot
and its light curves features. In section~\ref{model} we present a brief
description of the hydrodynamical model developed by \citet{mendoza09}
and the system of dimensionless units in which it is useful to
make comparisons with observations.  The fits to the light curves
using  the hydrodynamical model of \citet{mendoza09} are developed
in the section~\ref{fit}. The result of our fits and a discussion
of the obtained physical parameters of the model are presented in
section~\ref{discussion}.

\section[Observations]{Observational Data} 
\label{observations}

\begin{figure}
\begin{center}
 \includegraphics[scale=0.65]{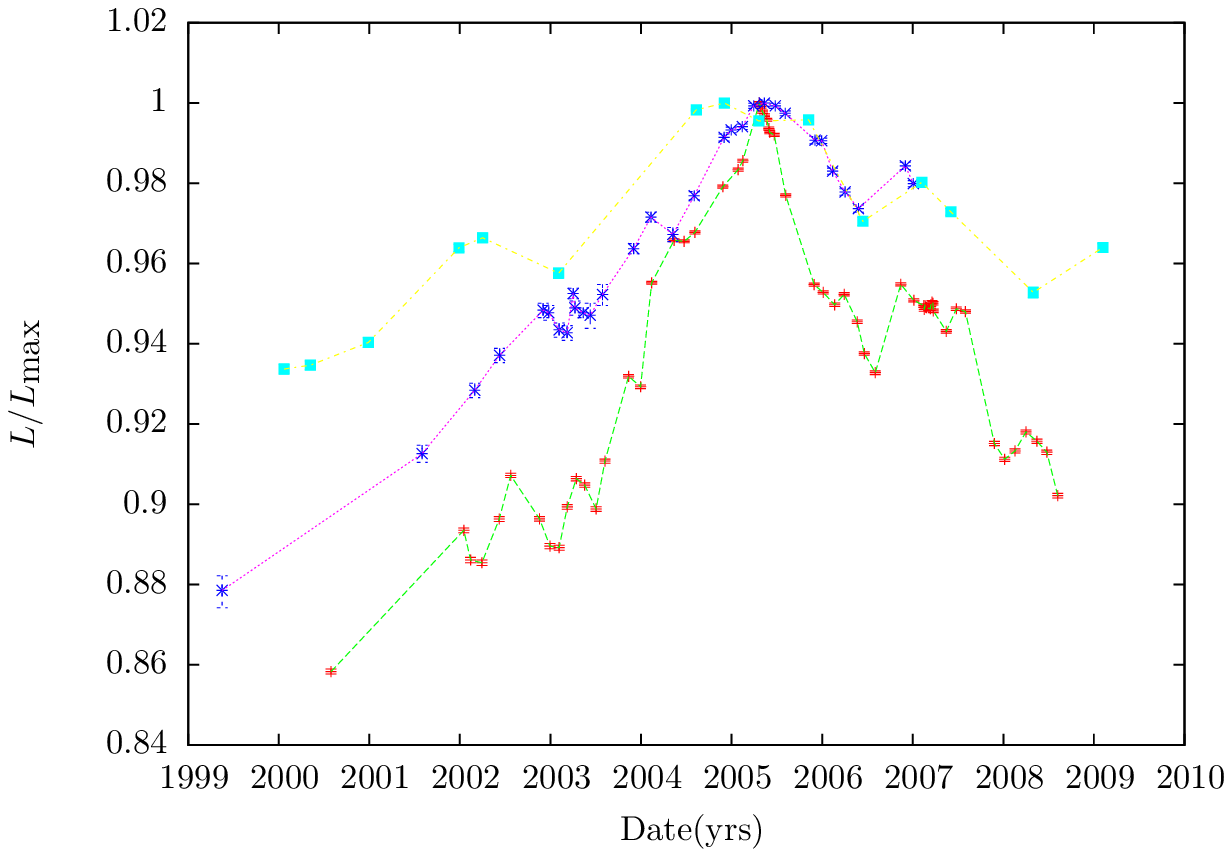}
\end{center}  
\caption{The figure shows multi-frequency luminosity curves of the 
HST-1 knot of the galaxy M87.  All curves have been normalised to the
maximum.  From bottom to top, the curves represent \( 2 \text{cm}
\) radio \citep{Chang10}, \( 225.5 \text{nm} \) UV \citet{madrid09} and
Chandra \( 2 \text{Kev} \) X-ray \citep{harris09}.  }
\label{fig00}
\end{figure}

  The multi-frequency light curves were taken from three separate datasets
and are shown in Figure~\ref{fig00}.  
X-ray observations were taken from a multi-frequency program 
coordinating Chandra and HST monitoring  \citep{harris09}. 
Ultraviolet data are part of the same program and carried out 
during the years 1999 to 2006 \citep{madrid09}. Finally the radio 
data corresponds to observations with the VLBI at \( 2 \text{cm} \)
\citep{Chang10}.  As mentioned in section~\ref{introduction},  these all
show a clear outburst with a maximum emission occurring in 2005, followed
by a small outburst.  After this,
subsequent micro-outbursts differ from each other in the global decay of
the light curve.  Although all the observational data show the same morphology 
in the light curves, the spectral power law differs in each section 
of the spectrum from radio to X-rays, revealing that a simple power 
law cannot describe the whole spectra of HST-1 \citep{harris09}.

  We calculate the flux in X-rays following the procedure described
by \citet{harris06} and applied it  to the observational intensities
of the HST-1 knot reported by \citet{harris09}.  The flux in
ultraviolet and radio wavelengths is calculated with a conversion
factor between \(\textrm{Jy}\) to \(\textrm{W}/\textrm{m}^2\), using
a reference wavelength of \(225.55\textrm{nm}\) for the ultraviolet
data \citep{madrid09} and \(2\textrm{cm}\) for the radio measurements
\citep{Chang10}.

  We assume a negligible extinction factor and an isotropic emission of
the source, located at a distance of \(16 \textrm{Mpc}\) corresponding to
the distance to the galaxy M87 \citep{Jordan05}. With these assumptions
we obtain a lower limit for the luminosity of the HST-1 knot in different
wavelengths.

\section[model]{Model}
\label{model}

  Let us assume that periodic injections of velocity and mass flows are
injected at the base of a 1D relativistic flow moving along a jet and
consider a particular time on the ejection process in which a fast parcel
of flow is ejected after a slow one.  A time later, the fast parcel will
``overtake'' the slow one and the flow will become multi-valued.  In order
to arrange this contradiction, nature creates an initial discontinuity of
the hydrodynamical values which later develops into a working surface,
i.e. a contact discontinuity bounded by two shock waves,  that moves
along the jet on the direction of the flow as measured from the central
engine \citep[see e.g.][]{Landauhydro}.

  The first ideas about radiative internal shock waves inside an
astrophysical jet were developed by \citet{rees94,daigne98}.   
Although 
several extensions and particular aspects of the model have been 
presented in the literature \citep[see e.g][]
{spada99,spada01,misra05}. A semi-analytical description of this 
phenomenon was made by M09.  These last model assumes periodic injections of
mass and velocity at the base of the jet.  Using mass and 
conservations of the ejected material, it is possible to 
account for the kinetic power loss as the working surface travels 
along the jet, assuming that the radiation time scales are small 
compared to the characteristic dynamical times of the problem.
The pressure of the fluid is thus negligible and 
so the description of the flow can be well described by a ballistic
approximation. This assumption is 
valid if the flow within the jet is nearly adiabatic and non--turbulent 
\citep[see e.g.][]{misra05}. In what follows we will use the model by 
M09 to describe the multi-wavelength light curve features of the HST-1 
knot in M87.

  To follow the evolution of the working surfaces M09 considered
a source ejecting material in a preferred direction with a velocity
\(v(\tau)\) and a mass ejection rate \( \dot{m}(\tau) \), both dependent
on the time \( \tau \) as measured from the jet's source. A further
assumption is made such that the working surface is thin and mass losses
within it are negligible. The energy loss \(E_\text{r}\) by the working
surface is given by \(E_\text{r} = E_{0}- E_\text{ws}\), where \(E_{0}\) is the
injected energy at the base of the jet and \(E_\text{ws}\) is the energy
inside the working surface. The kinetic power  available within the working
surface is then given by \( \mathrm{d}E_{r}/ \mathrm{d} t\).  If this
power is converted efficiently into radiated energy then the Luminosity
\( L \) produced by the emission of the working surface is given by \(
L = - \mathrm{d}E_{r}/ \mathrm{d} t \).

  On the one hand, we assume that the injected  velocity at the base of the jet 
is a periodic function of time, given by:

\begin{equation}
  v(\tau)  = v_{0} + c \eta^2 \sin \left( \omega\tau \right),
\label{eq20} 
\end{equation}

\noindent where the velocity \(v_{0}\) is the ``background'' average bulk
velocity of the flow inside the jet and \(\omega\) is the oscillation
frequency of the injected velocity. The positive dimensionless parameter
\(\eta^{2}\) measures the amplitude variations of the flow and is such
that the oscillations of the flow are sufficiently small, in such a way
that the total bulk velocity \(v(\tau)\) does not exceeds the velocity
of light \( c \).

  On the other hand, the mass ejection rate \(\dot{m}\) 
injected at the base of the jet has the following periodic variation:

\begin{equation}
  \dot{m}  = \dot{m}_0 + \dot{\mu} \sin \left( \Omega \tau \right).
\label{eq30} 
\end{equation}	

\noindent where \( \dot{m}_0 \) is the  ``background'' average mass
ejection rate and \( \Omega \) is the oscillation frequency of the mass
ejection rate.  The parameter \( \dot{\mu} \) is the amplitude of the injected
oscillation.

  In the original article by M09 and in further applications
\citep[see e.g.][]{cabrera13,coronado14} the  modelling of outbursts for
long gamma-ray bursts, blazars and micro-quasars was performed under the
assumption that \(
\dot{\mu}  = 0 \) and so \( \dot{m} = \text{ const. } \)  Although this
simplifies the number of free parameters of the model, it turns out that
the light curve of the HST-1 knot in M87 cannot be modelled with such a
simple assumption.

  In order to use the semi-analytical ballistic M09 model on its more
general form, we proceed as follows.  The model depends on six unknown
parameters: \( v_0,\ \eta^2,\ \omega,\ \dot{m}_0,\ \dot{\mu},\ \text{and
} \Omega \).  To reduce the number of unknown parameters, we proceed
as follows.

  The luminosity \( L \) depends on six dimensional parameters: \(
v_0,\ c \eta^2,\ \omega,\ \dot{m}_0,\ \dot{\mu},\ \text{and } \Omega \).
Additionally, the velocity of light \( c \) is an important dimensional
parameter of the relativistic phenomena we are dealing with and so,
it has to be added to the list of important dimensional quantities of
the problem.  Since there are three fundamental independent dimensions,
namely the dimensions of time, length and mass, Buckingham's \( \Pi
\)-Theorem of dimensional analysis means that the luminosity can be
described as follows:

\begin{equation}
  L = \dot{m}_0 c^2 \, L'\left(v_0/c,\ \eta^2,\ \dot{\mu} / \dot{m}_0,\ 
         \Omega / \omega \right).
\label{s01}
\end{equation}

\noindent In the previous equation, the dimensionless luminosity \( L' \)
is a function of the four dimensionless quantities \( v_0/c,\ \eta^2,\ 
\dot{\mu} / \dot{m}_0,\ \Omega / \omega \).  In other words, the seven
dimensional quantities for which the luminosity depends on, can be reduced
to the problem of only four dimensionless quantities.  

\section{Fits to the observational data}
\label{fit}

  The observed and theoretical luminosities, \( L_\text{obs} \) and \(
L_\text{th} \) respectively, can be fit to the observational data with
the use of their dimensionless counterparts \( L_\text{obs}' \) and \(
L_\text{th}' \) by rescaling them as follows.   Both theoretical and
observed dimensionless luminosities can be normalised to their maximum
values: \( L'_\text{obs} (\tau'_\text{max}) \) and \( L'_\text{th}
(\tau'_\text{max}) \), i.e.

\begin{equation}
  \mathbb{L}_\text{obs} := \frac{ L'_\text{obs}  }{
  L'_\text{obs}(\tau'_\text{obs,max}) }, \qquad \quad \mathbb{L}_\text{th} := 
     \frac{ L'_\text{th}  }{ L'_\text{th} (\tau'_\text{th,max}) },
\label{s02}
\end{equation}

\noindent where  the dimensionless times \( \tau'_\text{obs,max}  \) 
and \( \tau'_\text{th,max}\) correspond
to the particular times where the observed or theoretical luminosities 
reach a maximum value respectively. According to Buckingham's \( \Pi
\)-Theorem of dimensional analysis, the dimensionless time \( \tau'
\) is related to the time \( \tau \) by the following relation:

\begin{equation} 
  \tau = \omega^{-1} \, \tau'.
\label{s04}
\end{equation}

\noindent In order to measure the observed and theoretical times in the
same system of dimensionless units we normalised them to the time given by
the FWHM of the outburst, i.e.:

\begin{equation}
  \mathbb{T}_\text{obs} :=  \frac{ \tau'_\text{obs} }{
     \tau'_\text{obs}(\text{FWHM}) }, \qquad \quad  
   \mathbb{T}_\text{th} :=  \frac{ \tau'_\text{th} }{
     \tau'_\text{th}(\text{FWHM}) }.
\label{s03}
\end{equation}

  The best fit of the theoretical luminosity \( \mathbb{L}_\text{th}
( \mathbb{T}_\text{th}) \) to the observed light curve \(
\mathbb{L}_\text{obs} (\mathbb{T}_\text{obs} ) \) yields a direct best
value for the four dimensionless free parameters \( v_0/c,\ \eta^2,\
\dot{\mu} / \dot{m}_0,\ \Omega / \omega \).  
The quantity \( \dot{m}_0 \) is obtained by using~\eqref{s01} evaluated
at one particular point of the light curve, which we choose as the point
where the light curve reaches its maximum value.  Once this last quantity
is known, the value for the parameter \( \dot{\mu} \) is hence inferred.
The frequency \( \omega \) is obtained using equation~\eqref{s04}
evaluated at a particular time, which we choose as the time where the
light curve reaches its maximum value.  With this, the parameter \(
\Omega \) is then inferred.

  The parameter calibration of the model is conceptualised as an
optimisation problem and so, we propose to solve it using Genetic
Algorithms (GAs), which are evolutionary based stochastic search
algorithms that mimics natural evolution. In this heuristic search
technique, points in the search space are considered as individuals
(solution candidates), which as a whole form a population. The particular
fitness of an individual is a number, indicating its quality for the
problem at hand. As in nature, GAs include a set of fundamental genetic
operations that work on the genotype, i.e. the solution candidate
codification, namely: mutation, recombination and selection operators
\cite{Mitchell-98}. These algorithms operate with a population of
individuals $P(t) = {x_{1}^{t}, ...,x_{N}^{t}}$, for a particular $t$
iteration, where the fitness of each $x_{i}$ individual is evaluated
according to a set of objective functions $f_{j}(x_{i})$. This objectives
functions allows to order from best to worst individuals of the population
in a continuum of degrees of adaptation.  Individuals with higher fitness,
recombine their genotypes to form the gene pool of the next generation,
in which random mutations are also introduced to produce a new variability.

  A fundamental advantage of GAs versus traditional methods is that
GAs solve discrete, non-convex, discontinuous, and non-smooth problems
successfully and so, they have been widely used in Ecology, Natural
Resources Management, among other fields \citep{Lopez-Corona-13} with
some astrophysical applications \citep[see e.g.][]{statbook}.  Our GA
evaluated the luminosity function  \( \mathbb{L}_\text{th}\left(v_0/c,\
\eta^2,\ \dot{\mu} / \dot{m}_0,\ \Omega / \omega \right) \)
of M09 in order to compare numerical
results from the model with the observed light curve \(
\mathbb{L}_\text{obs} \)  using standard Residual Sum
of Squares (RSS) as objective functions.  All parameters were searched in
the broadest possible range: \( 0.1 \lesssim v_0 / c \lesssim 0.999 \), \(
0.0001 \lesssim  \eta^2   \lesssim 0.899 \),  \( 0.001 \lesssim \dot{\mu} /
\dot{m}_0 \lesssim 1.0 \) and \( 0.001 \lesssim \Omega / \omega \lesssim 20
\).  The choice is consistent with the physical restriction of keeping
subliminal the full bulk velocity of the flow \( v \) and to the fact that
a large value of \( \dot{\mu}/\dot{m}_0  \)  would yield a huge  unphysical 
luminosity value.  A very large value of \( \Omega / \omega \) produces
large mass ejection oscillations, something not clearly visible from the
light curves.  This search parameter technique 
generates populations of 100 possible
solutions over a maximum 5000 generation search process, with a total
of 500000 individuals.  The GA algorithms selected were: tournament
selection with replacement \citep{goldberg1989messy,sastry2001modeling},
simulated binary crossover (SBX) \citep{deb2} and  polynomial mutation
\citep{deb2,deb2001multi}. The obtained final parameters were estimated
by averaging the 500 best individuals.  

\begin{figure}
\begin{center}
 \includegraphics[scale=0.65]{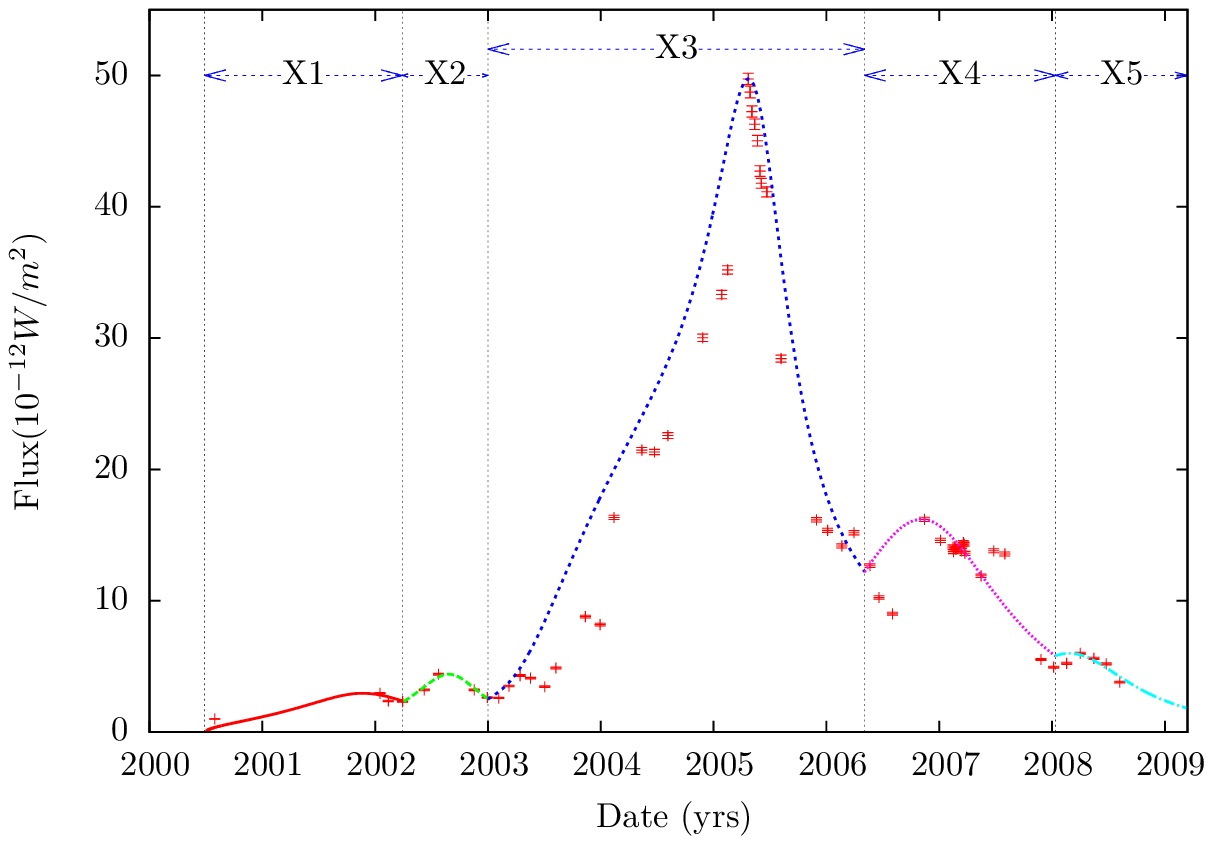}
\end{center}  
\caption{The figure show the fits (lines) to the X-ray data points of the 
light curve of the HST-1 knot observed by \citet{harris09} using the model
by M09.  The data points were divided into 5 time sections marked by the
dotted vertical lines corresponding to individual outbursts, and 
labelled X1, X2, X3, X4 and X5. The resulting calibration of the free
parameters of the model by M09 to the observed light curve are shown in
Table~\ref{tabla}.
}
\label{fig05}
\end{figure}

\begin{figure}
\begin{center}
 \includegraphics[scale=0.62]{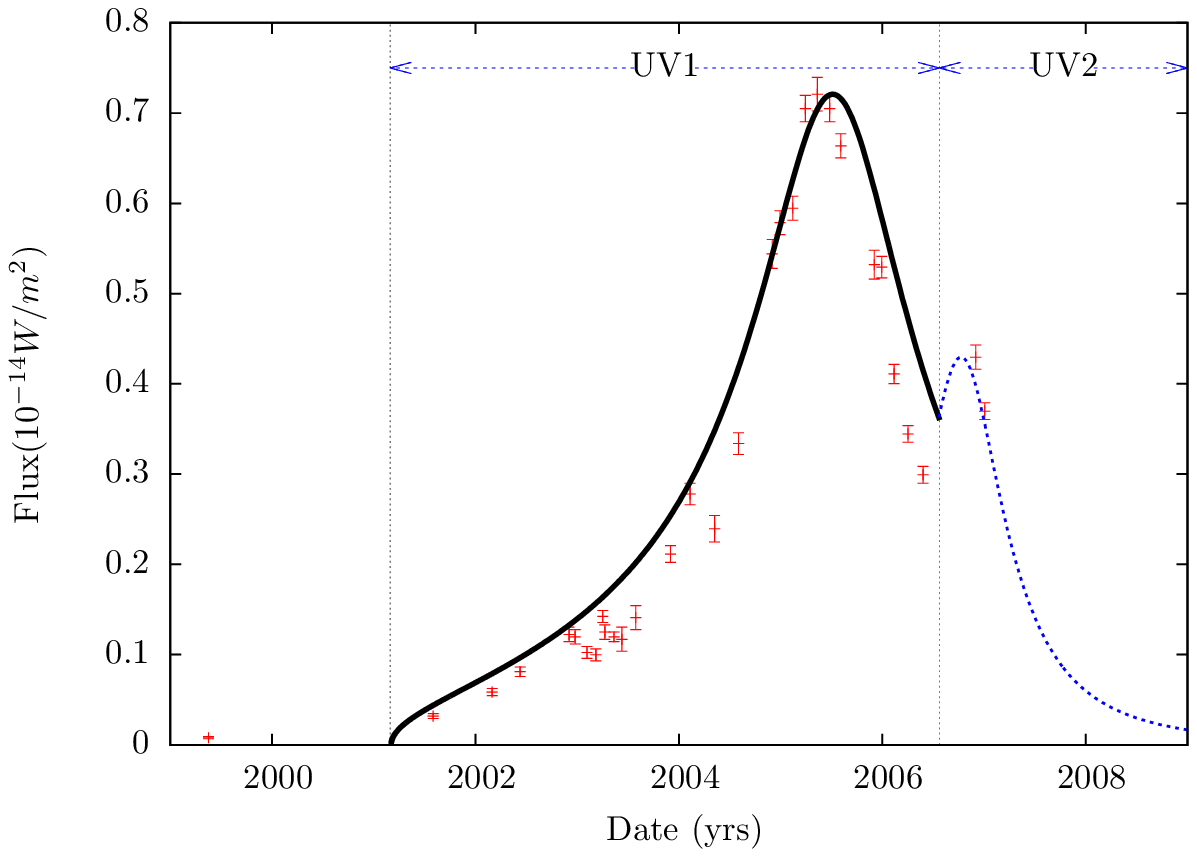}
\end{center}  
\caption{The figure show fits (lines) to the UV data points of the light
curve of the HST-1 knot observed by \citet{madrid09} using the model
by M09.  The data points were divided into 2 time sections marked by the
dotted vertical lines corresponding to individual outbursts, and 
labelled UV1 and UV2. The resulting calibration of the free
parameters of the model by M09 to the observed light curve are shown in
Table~\ref{tabla}.
}
\label{fig06}
\end{figure}

\begin{figure}
\begin{center}
 \includegraphics[scale=0.65]{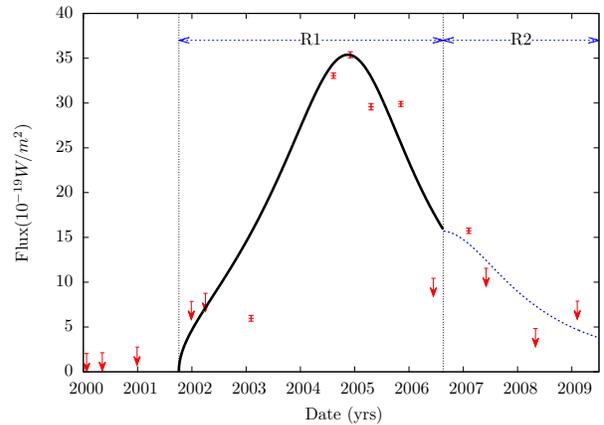}
\end{center}  
\caption{The figure show fits (lines) to the radio data points of the light
curve of the HST-1 knot observed by \citet{Chang10} using the model
by M09.  The data points were divided into 2 time sections marked by the
dotted vertical lines corresponding to individual outbursts, and 
labelled R1 and R2. The resulting calibration of the free
parameters of the model by M09 to the observed light curve are shown in
Table~\ref{tabla}.
}

\label{fig07}
\end{figure}

  Direct inspection of the light curves in Figure~\ref{fig00} show that
multiple outbursts occur during the period of observation.  As such and
following the procedure of \citet{cabrera13} we divided the light curves
into individual outbursts.  Two clear outbursts appear on all wavelength
observations and an additional 3 mini-outburst were defined for the X-ray
data -two before the main outburst and one at the end of the observations.
The results of the GA explained above for each outburst are presented 
in Table~\ref{tabla} and the best fits
to the light curves with these parameters are shown in 
Figures~\ref{fig05}-\ref{fig07}.


\begin{table*}
 
\begin{tabular}{@{}lcccccccccc}
\hline 
ID & $v_{0}/c$ & $\eta^{2}$ & $\dot{m}_{0}$ & $\dot{\mu}$ & $\dot{m}_{\textrm{max}}$ & $\omega_{0}^{-1}$ & $\Omega^{-1}$ & $\Gamma_{\textrm{min}}$ & $\Gamma_{\textrm{max}}$ & $\Gamma_{\textrm{bulk}}$\tabularnewline
 &  &  & $(10^{-3} M_{\odot}\textrm{yr}^{-1})$ & $(10^{-3}
 M_{\odot}\textrm{yr}^{-1})$ & $(10^{-3} M_{\odot}\textrm{yr}^{-1})$ &
 $(\textrm{days})$ & $(\textrm{days})$ &  &  & \tabularnewline
\hline 
\hline 
$\textrm{X}_{1}$ & 0.9631  & 0.0360  & 4.252 & 0.415 & 4.667 & 53.8 & 0.716 & 2.67 & 23.12 & 3.72\tabularnewline
\hline 
$\textrm{X}_{2}$ & 0.9573 & 0.0420 & 3.476 & 1.119 & 4.596 & 26.4 & 4.361 & 2.48 & 27.32 & 3.46\tabularnewline
\hline 
$\textrm{X}_{3}$ & 0.8156 & 0.1839 & 9.727 & 3.943 & 13.671 & 374 & 2322.5 & 1.29 & 31.32 & 1.73\tabularnewline
\hline 
$\textrm{X}_{4}$ & 0.9713 & 0.0275  & 36.56 & 3.351 & 39.916 & 118 & 45.874 & 3.03 & 20.16 & 4.2\tabularnewline
\hline 
$\textrm{X}_{5}$ & 0.9150 & 0.0823 & 4.511 & 2.057 & 6.569 & 83.5 & 36.0 & 1.81 & 13.49 & 2.48\tabularnewline
\hline 
 &  &  & $( 10^{-6} M_{\odot}\textrm{yr}^{-1} ) $ & $( 10^{-6}
 M_{\odot}\textrm{yr}^{-1}) $ & $( 10^{-6} M_{\odot}\textrm{yr}^{-1} ) $ &  &  &  &  & \tabularnewline
\hline 
\hline 
$\textrm{UV}_{1}$ & 0.9020 & 0.0976  & 1.569 & 0.588 & 2.157 & 296 & 50.818 & 1.68 & 32.97 & 2.32\tabularnewline
\hline 
$\textrm{UV}_{2}$ & 0.9762 & 0.0228 & 8.094 & 3.338 & 11.433 & 16.1 & 4.139 & 3.31 & 22.83 & 4.61\tabularnewline
\hline 
\hline 
 &  &  & $( 10^{-9} M_{\odot}\textrm{yr}^{-1} ) $ & $( 10^{-9}
 M_{\odot}\textrm{yr}^{-1} ) $ & $ ( 10^{-9} M_{\odot}\textrm{yr}^{-1} ) $ &  &  &  &  & \tabularnewline
\hline 
\hline 
$\textrm{R}_{1}$ & 0.9450 & 0.0536 & 2.650 & 2.033 & 4.683 & 195.5 & 33.072 & 2.21 & 19.18 & 3.06\tabularnewline
\hline 
$\textrm{R}_{2}$ & 0.9724 & 0.0270 & 9.156 & 2.821 & 11.978 & 75.43 & 31.21 & 3.07 & 27.12 & 4.28\tabularnewline
\hline 
\end{tabular}
\medskip

\caption{Best parameter estimations  using the X-ray, 
UV and radio light curves of the HST-1 knot of the galaxy M87.  The
resulting light curves are shown in Figures~\ref{fig05}-\ref{fig07}.  The
fits were performed by dividing the light curves in time ID sections 
represented by the first column of the table.   The quantity \(
\dot{m}_\text{max} \) corresponds to the maximum mass ejection rate
discharged by the jet for a particular outburst.  The values \(
\Gamma_\text{min} \), \( \Gamma_\text{max} \) and \( \Gamma_\text{bulk} \)
are the minimum, maximum and background (i.e. \( v_0 \) bulk ``average''
velocity of the flow) Lorentz factors of the flow.  All parameters were
obtained to a precision above a 2-\(\sigma\) statistical confidence level.
}
\label{tabla}
\end{table*}


\section{Discussion}
\label{discussion}

  Every modelling process goes an initial exploratory face in which a basic  
hypothesis space is set up.  In this context, \citet{williams14} found that
a good modelling process should: (a) stay as close as data as possible, (b)
includes as much phenomenological information as possible and (c) keep as
simple as possible.  
  
  The parameter estimation of the model is quite close to the observational
data (since it has a \( \gtrsim 2\sigma \) confidence level value), with a
simple ballistic model describing a complicated hydrodynamical phenomenon. 

  At first sight, the curves seem not to properly adjust to many data
points, as one should expect with such small observational uncertainties
in the data.  However, the time series represented by the light curve
has many temporal gaps. Between these temporal gaps, the value of the
inferred physical parameters  may not stay the same, making the light
curve to present mini-outbursts combined with different oscillations.
For example, the data points about 2007 in X-rays can be modelled as
a series of mini-outbursts.  But modelling such a number of mini-bursts
in a context of insufficient physical data represents an increase of
unjustified additional hypothesis, despite the fact of an increment
in  statistical accuracy.  As pointed out by \citet{ross00} it should
be expected some sort of conflict between parsimony and realism.
Nevertheless as models tends to incorporate more hypothesis, and become
increasingly complex there is lose in transparency interpretation.

  Although equation~\eqref{s01} is dimensionally correct it doesn't
take into account the fact that an efficiency proportionality factor \(
\xi \) should appear in the right hand side of the equation, i.e. \( L =
\xi \dot{m}_0 c^2 L' \).  This factor does not only depend on the ratio of
the radiated luminosity to kinetic loss power inside the working surface,
but also on the frequency of the emitted radiation.  This is the reason
as to why the inferred mass ejection rates for the same outburst differ
so much at different wave-lengths.  In other words, at best one should
consider the values of the mass ejection rates in Table 1 as lower limits.
The inferred  Lorentz factors for the bulk flow are \( \sim 1 - 4 \)
reaching maximum values of up to \( \sim 30 \).

  The model by M09 has shown to be quite useful reproducing light
curves of long gamma-ray bursts, blazars and micro-quasars.   As we have
shown in this article, the same model is also good for dealing with the
light curve of the HST-1 knot of M87.  Our modelling can be adjust more
precisely to the observed data by suitably performing more subdivisions
of the data set, essentially modelling many mini-outburst.  Since no
data is available for these mini-outbursts, their introduction would
be speculative.  In this sense, the current modelling can be interpreted
as a baseline modelling \citep{schwab13} that captures the key patterns
in the empirical data and the associated physical processes.

\section*{Acknowledgments} This work was supported by DGAPA-UNAM
(IN111513-3) and CONACyT (240512) grants. YC, OL and SM acknowledge
economic support from CONACyT (210965, 62929 and 26344).  OL  acknowledges
economic support from a DGAPA-UNAM fellowship.

\bibliographystyle{mn2e}
\bibliography{lightcurves}

\end{document}